\documentclass[%
 reprint,
 superscriptaddress,
 showpacs,
  nofootinbib,
 amsmath,amssymb,
 aps,
   prl,
]{revtex4-1}

\usepackage{verbatim}
\usepackage{graphicx}
\usepackage{epsfig}
\usepackage{amsfonts}
\usepackage{amsmath,amssymb}
\usepackage{bm}
\usepackage{hyperref}
\usepackage{color}

\begin{document}

\title{Nucleon-nucleon resonances at intermediate energies using a complex energy formalism.}

\author{G. Papadimitriou}
\email{georgios@iastate.edu}
\affiliation{
Department of Physics and Astronomy, Iowa State University, Ames, Iowa 50011-3160, USA
}%

\author{J. P. Vary}
\email{jvary@iastate.edu}
\affiliation{
Department of Physics and Astronomy, Iowa State University, Ames, Iowa 50011-3160, USA
}%


\begin{abstract}

We apply our method of  complex scaling, valid for a general class of potentials, 
in a search for nucleon-nucleon S-matrix poles up to 2 GeV laboratory kinetic energy. 
We find that the realistic potentials 
JISP16, constructed from inverse scattering, and chiral field theory potentials 
N$^3$LO and N$^2$LO$_{opt}$ support resonances in energy regions well above their fit regions. 
In some cases these resonances 
have widths that are narrow when compared with the real part of the S-matrix pole.

\end{abstract}


\maketitle



{\it Introduction.} 
There is a long-standing interest and considerable recent progress in the theoretical characterization of nuclear resonant states.
A resonant state is fully characterized by its position in the energy plane and its width, which determines how fast the
state will decay.  One could in general solve the time-dependent Schr\"odinger  equation
to study the characteristics of  resonant states \cite{Talou,Hagino,Volya}, which is a demanding computational process.
On the other hand, the time-independent many-body methods that deal with the description of resonant states in nuclei are under development and exhibit appealing computational features.  These time-independent methods  can be divided into real energy and 
complex energy approaches. 

The spectrum of a real nuclear Hamiltonian consists of negative and positive energy states. While the negative energy spectrum
is discrete (bound states), the positive energy spectrum may have  a richer structure with resonant states among scattering or continuum states.
Hence real-energy approaches require criteria for identifying a resonant structure and for assigning a position and a width to them. 

In the domain of $\mathcal{L}$$^2$ integrable basis expansion methods this is usually achieved through  $\mathcal{L}$$^2$
stabilization methods \cite{hazi,Salzgeber} or methods that evaluate the real Continuum Level Density (CLD) \cite{Shlomo}.
The CLD usually produces an approximate 
Breit-Wigner distribution in the region of the resonant state whose parameters could be determined by a fit. The CLD method has been
successfully applied to atomic systems \cite{Arai_kruppa}, nuclear clusters \cite{Arai_kruppa2} and in mean-field approaches for 
describing quasiparticle resonant states \cite{Pei}. 

The name stabilization, arises from the fact that one does not need the knowledge of the asymptotic wavefunction in order to determine the resonant parameters.
On the reaction side, based on R-matrix considerations \cite{arima,Lane,Descouvemont}, and assuming the single channel approximation, resonant
parameters can be determined by the behavior of the phase-shift function of energy $\delta$(E) around the resonant position; in particular
the position is defined as the inflection point of $\delta$(E) (maximum energy derivative of $\delta$(E)) and the width is defined as $\Gamma$ = $\frac{2}{d\delta / dE}|_{E=E_r}$,
where $E_r$ is the inflection point. Such formulas
where employed recently in microscopic $R$-matrix calculations \cite{Navratil1} to extract widths from realistic nucleon-nucleus phase-shifts.
Though the R-matrix parametrizations have been very successful, the formulas become less transparent in the multi-channel case and  when they are applied for the description of broad resonances (see discussion in \cite{Thompson}).
Furthermore, for broad resonances, R-matrix analyses become more dependent on 
  channel radii  and boundary conditions   
  \cite{Hale_Csoto}. 
Finally, combining formulas and assumptions from different theories/models for the calculation of an observable increases the possibility of uncontrollable errors.

The complex energy formalism 
serves as a potentially fruitful alternative for the characterization of the resonant parameters.
It has been shown that once the R-matrix, S-matrix and T-matrix are analytically continued to the complex energy plane, the extraction of resonant 
parameters becomes independent of boundaries and radii \cite{Compiled_data}. 
Apart from this practical issue, some physical phenomena may have 
a more natural interpretation once the theory is developed in the complex energy plane (e.g. thermo-nuclear reactions \cite{jarmie,lovas}).
In the complex energy formalism, the Gamow (resonant) states,  i.e. the solutions of the Schr\"odinger equation  
which satisfy purely outgoing boundary conditions (complex wave number $k$), play a dominant role. 
It was shown by Berggren \cite{Berggren} that resonant states, when accompanied
by non-resonant continuum states, form a complete set, an important property that gives rise  to Berggren basis expansion methods either in a Configuration Interaction (CI)
framework \cite{Michel_2002,Betan_2002,Rotureau_2006,review_gsm,betan_witek,ncgsm}, Coupled Cluster framework \cite{Hagen_helium,gaute_Oxygen,gaute_ca48} or 
reaction theory framework \cite{gamow_rmatrix,gamow_rmatrix2,Jaganathen,gaute_michel,betan_2014,fossez1,fossez2}.
Expressing the Hamiltonian in such a complex energy, orthonormal non $\mathcal{L}^2$ integrable basis, automatically allows its spectrum to support resonant and also non-resonant continuum states. In addition, when the Berggren basis is used in a reaction framework the detailed knowledge of the boundary condition at large distances is not crucial.

The Complex Scaling (CS) method also belongs in the category of complex energy formalisms.
The Aguilar-Balslev-Combes (ABC) theorem \cite{ABC1,ABC2} establishes that once the Hamiltonian coordinates are rotated, the resonant states are independent of the rotation and behave 
asymptotically as bound states. Consequently, one could use the technology that has been established for bound states in order to  describe resonant and scattering phenomena. Furthermore, the CS method has been successfully 
applied in nuclear physics \cite{Kruppa_1986,Myo20141,lazauskas1,Glockle,kruppa_george} 
(see also \cite{liu_quan} for an application of CS in a deformed nuclear mean-field). 
We recently showed \cite{cs_real2} that this method may be applied to the most general cases of non-local nuclear potentials.

In this work we apply the CS method to nucleon-nucleon (NN) scattering spanning the range from threshold to 2 GeV laboratory kinetic energy, which exceeds the fitting range of most NN potentials.  We elect to retain non-relativistic kinematics throughout as the interactions are derived for a non-relativistic scattering framework. We employ three different realistic NN interactions and we find resonant poles at laboratory kinetic energy of about 600 MeV, or at about 2.2 GeV in the total center of mass energy. Some of these poles correspond to narrow resonant states.

According to the SAID data analysis group \cite{arndt1,arndt2,arndt3} (see also \cite{Strakovsky}) there 
exist resonance-like structures, poles of the S-matrix, in the $^1$D$_2$, $^3$F$_3$ uncoupled and coupled $^3$P$_2$-$^3$F$_2$ channels.
 Recently a resonant-like structure was also found by the WASA-at-COSY collaboration  and the SAID analysis group 
 in the $^3$D$_3$-$^3$G$_3$ coupled channel \cite{Adlarson1,Adlarson2}. 
 Our CS  calculations, in addition to showing resonances in these channels, also reveal resonance-like structures 
 in the $^3$P$_0$ and $^3$P$_1$ channels. We 
searched other channels
up to and including L=4 without any additional signals of resonance-like structures.

The study of dibaryon resonances could shed light on the reaction mechanism and aid in the 
interpretation of excited nucleonic states. 
It is also valuable to pin down the properties of dibaryon resonances as a potential link between Quantum Chromodynamics, hadron models and traditional low energy nuclear physics.
In the work of \cite{arndt1,arndt2,arndt3} the resonant-like structures where identified by analytically continuing the T-matrix of the available data in the
complex energy plane. 
Our goal is to simply identify resonant structures with the CS method but not to study in depth the characteristics of the NN scattering at intermediate energies, something that
would require the use of NN interactions that fit scattering data at higher energies, such as CDBonn \cite{cdbonn} or AV18 \cite{av18}.
Such in-depth studies would be done relativisticly \cite{elster} and by treating properly $\Delta$ and/or Roper resonances degrees of freedom (see for example \cite{Epelbaum}).
Furthermore, we will not provide  information on the possible decay paths that the resonant structures will follow, since we 
do not consider couplings to inelastic channels such as, NN$\, \to \, \pi $d or NN$ \to $ $\Delta$N etc. In addition, the interactions we use,
are modeling the short-range (high-energies) NN sector in different approaches  and are fitted at lower laboratory energies 
($\leq$ 350 MeV).
Hence, we are  
not aiming at making predictions for the existence or absence
of broad dibaryonic states. Nevertheless, it is worthwhile to discover that the NN interactions we employ, support high energy resonant-like states above the
$\Delta$ production threshold (1232 MeV). The existence of these resonances indicates such degrees of freedom 
suggest it may be important to examine what effect the explicit inclusion of the related degrees of freedom 
will have on finite nuclei.
Simply stated, our goal at this point is to demonstrate that the CS method
locates these resonances using three different realistic NN interactions 
in the conventional non-relativistic framework.

{\it Method and results}.
We apply the CS transformation to our Hamiltonian which consists of the relative kinetic energy $T$ and the realistic NN interaction $V$
 between the nucleons.
The complex rotated Hamiltonian has the form:
\begin{equation}
\label{eq_Hami}
H(r,\theta) = e^{-2i\theta}T + V(re^{i\theta}),
\end{equation}
where $\theta$ is the real CS rotation parameter. 
Nuclear potentials  are usually non-local in coordinate space but, for simplicity and compactness, the potential is taken 
to be local in Eq.\eqref{eq_Hami}. In \cite{cs_real2} we  presented our methodology (see also \cite{contour_dm}) for applying the 
CS transformation to a non-local potential which
we also follow here.
\begin{figure}[h!] 
  \includegraphics[width=\columnwidth]{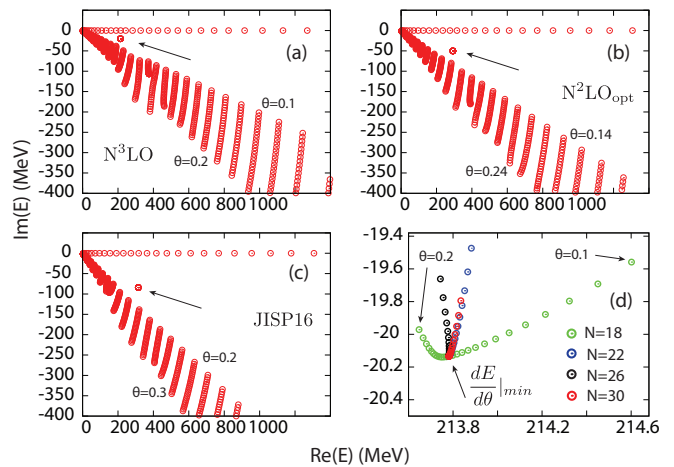} 
  \caption[T]{\label{Fig1}
  (Color online) Spectrum of the CS Hamiltonian \eqref{eq_Hami}  for  the N$^3$LO (a), N$^2$LO$_{opt}$ (b) and JISP16 realistic potentials for
  the $^3$P$_1$ channel (or 1$^-$ state)  of the $np$ system. There is a state that is invariant of the rotation angle $\theta$ (rad), indicated by an arrow. 
In panel (d) we show a magnification of the N$^3$LO Hamiltonian spectrum in the region of the resonant-like state at a sequence of cutoffs in the number of radial nodes (N) in the basis. This is the $\theta$-trajectory. Energies are in the 
Center of Mass (CoM) frame.}
\end{figure}
The time-independent non-relativistic Schr\"odinger equation then becomes:
\begin{equation}
\label{eq_eigval_cs}
H(r,\theta)\Psi(r,\theta) = E(\theta) \Psi(r,\theta).
\end{equation}
where E is the energy in the Center of Mass (CoM) frame here and throughout this work. To be more precise, the rotated non-Hermitian Hamiltonian operator and the Hermitian one, are related through the formula:
\begin{equation}
\label{rot_trans}
H(r,\theta) = U(\theta)H(r)U(\theta)^{-1},
\end{equation} 
where ${U}$$(\theta)$ stands for the non-Unitary CS transformation operator. It is then expected that  any
quantum mechanical operator ${O}$ will be transformed under \eqref{rot_trans} in the case of CS (e.g. see \cite{Horiuchi} for an application on the dipole operator). 
In order to calculate its expectation value, 
one could either use the transformed operator and calculate its expectation value between CS solutions ($\theta \neq 0.0$), or evaluate the
bare operator and calculate its expectation value using the coefficients from the back-rotated, $\theta$-independent solution. In \cite{kruppa_george} it
was shown that both ways are equivalent and, in addition, the back-rotation transformation was stabilized via a smoothing process.

In order to solve Eq.\eqref{eq_eigval_cs} we assume that the solution is a linear combination of orthonormal Harmonic Oscillator (HO) basis
states and we solve a complex symmetric Hamiltonian eigenvalue problem by diagonalization.
The spectrum of the Hamiltonian contains resonant (bound states, resonances) and non-resonant continuum states.  According
to the ABC theorem, once the resonant state is revealed it remains invariant under CS rotations, whereas the non-resonant 
continuum states follow an approximate 2$\theta$ path in the complex energy plane. This is the complex stabilization criterion that is used in CS
for the identification of the resonant state.  In practice, due to the truncation of the underlying basis, there is a small variation of the
resonant position with  $\theta$. 
It is then a consequence of the complex virial theorem \cite{virial_theom} that the resonant state will be the
one that corresponds to the minimal change of the real part of the energy with respect to $\theta$. The method is also known 
as $\theta$-trajectory method and it is a common practice in CS applications (see for example \cite{Aoyama,kruppa_george,masui}). We also apply this stabilization technique 
and we check convergence of our results as a function of the
basis dimension and the variations with the rotation angle.

In Fig.\ref{Fig1} we present the spectrum of the complex scaled Hamiltonian for the $^3$P$_1$ (1$^-$) channel  
of the neutron-proton ($np$) system, for the JISP16 \cite{jisp16} and the two
chiral effective field theory interactions N$^3$LO \cite{EM_pot} and N$^2$LO$_{opt}$ \cite{n2lo_opt}. 
The HO basis was characterized by $\hbar \omega$ = 40 MeV for JISP16 and N$^2$LO$_{opt}$ 
and by $\hbar \omega$ = 28 MeV for  N$^3$LO.
For the N$^3$LO potential we varied the rotation from $\theta$ = 0.1 rad to 0.2 rad, for N$^2$LO$_{opt}$ from $\theta$ = 0.14 to 0.24 rad
and for JISP16 from $\theta$ = 0.2 to 0.3 rad. The step in the $\theta$ discretization was 0.004 rad.
As we start rotating the coordinates and momenta of the Hamiltonian, solutions that initially inhabit the real-axis ($\theta$ = 0.0)
start moving inside the complex energy plane. During the rotation, when a state crosses the ordinates of a pole of the S-matrix, it remains there
and does not follow the rotation of the other non-resonant continuum states. 
In Fig.\ref{Fig1} we see clearly that  all interactions support one state which is almost invariant with respect to 
the CS rotation parameter. For the  calculations in panels (a), (b) and (c) we used a HO basis that consisted of a maximum of N=30 radial nodes.
The resonant parameters of this state where identified by applying the $\theta$-trajectory method. Any Hamiltonian after  
the CS transformation is applied, becomes non-Hermitian and one needs to locate the resonant state or the stationary point. 
The stationary point is the one for which the difference in energy with respect to the $\theta$ variation is minimal. The $\theta$-trajectory is 
\begin{figure}[h!] 
  \includegraphics[width=\columnwidth]{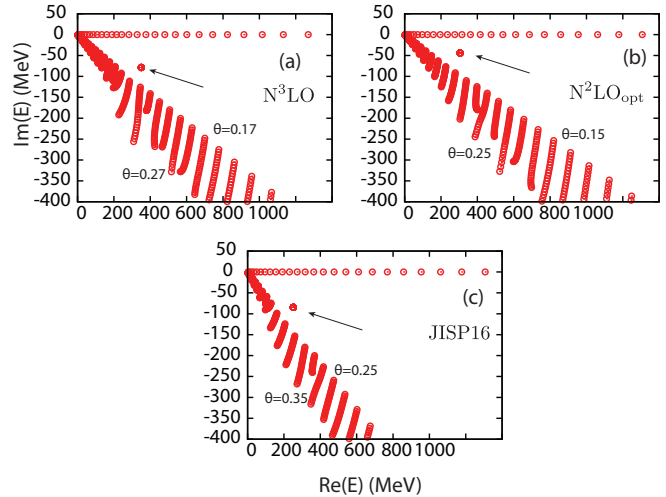} 
  \caption[T]{\label{Fig2}
  (Color online) Same as Fig.\ref{Fig1} but for the  $^3$P$_0$ channel (0$^-$ state) and for a HO basis size of N=30.}
\end{figure}
shown in panel (d) of Fig.\ref{Fig1}. We see that for each basis size (denoted by N) the $\theta$-trajectory is different. Nevertheless, the 
variations become smaller by increasing the basis size and the results for N=26 and N=30 start to coincide for $\theta$ $>$ 0.17 rad.
At this point we also notice that results are practically converged at N=18 and  differences appear only in the fifth and sixth significant digit.

In addition to the $\theta$-trajectory method,  the b-trajectory method is employed \cite{Aoyama} in CS calculations. 
This b-trajectory method uses the dependence of
the results on the length parameter of the basis, in our case the HO length or, equivalently, $\hbar\omega$. 
We do not analyze the b-trajectory method here. Here, we simply note that the deuteron 
ground state (g.s.) ($^3$S$_1$-$^3$D$_1$ coupled channel) is very well converged (independent of $\hbar \omega$) at
N = 30.  We will perform detailed b-trajectory analyses in future applications of CS to resonances in finite nuclei. 

We gather a sample of our results for the $^3$P$_1$ state in Table \ref{Tab:1}.
\begin{table}[ht]
\caption{$^3$P$_1$ resonant parameters in MeV as a function of the HO basis size. The choice of the energy was decided by the $\theta$-trajectory method. The energies  
       are in the CoM of the $np$ system above the $np$ threshold (E=0). To obtain the width, the formula 
       $\Gamma$ = -2Im(E) could be applied. For the $np$ system the nucleon  mass is taken to be 938.2 MeV/c$^2$.}
\centering
\begin{tabular}{c c c c}
\hline\hline
N & N$^3$LO & N$^2$LO$_{opt}$ & JISP16 \\ [0.5ex] 
\hline
14  & (214.767 -i20.048)  &   (292.114 -i49.609)  &  (314.471 -i84.250) \\
18  & (213.758 -i20.139)   &   (292.358 -i50.039)   &   (314.475 -i84.244) \\
22  & (213.784 -i20.131)  &   (292.323 -i50.023) &  (314.473 -i84.244) \\
26  & (213.781 -i20.134) & (292.327 -i50.020) & (314.473 -i84.244) \\
30  & (213.781 -i20.134) & (292.327 -i50.020) & (314.473 -i84.244) \\ [1ex]
\hline
\end{tabular}
\label{Tab:1}
\end{table}
The sensitivity analysis regarding the $\theta$ dependence and the basis convergence
was also performed for the other partial waves.

In Fig.\ref{Fig2} we present the CS spectrum for the $^3$P$_0$ channel. For a basis size of N=30  the converged resonant-like states are
(350.859 -i78.689) MeV,  (305.859 -i44.827) MeV and (251.252 -i84.454) MeV for the N$^3$LO, N$^2$LO$_{opt}$ and JISP16 interactions
respectively. 

We now focus our attention on the N$^3$LO and JISP16 interactions, since they
\begin{figure}[h!] 
  \includegraphics[width=\columnwidth]{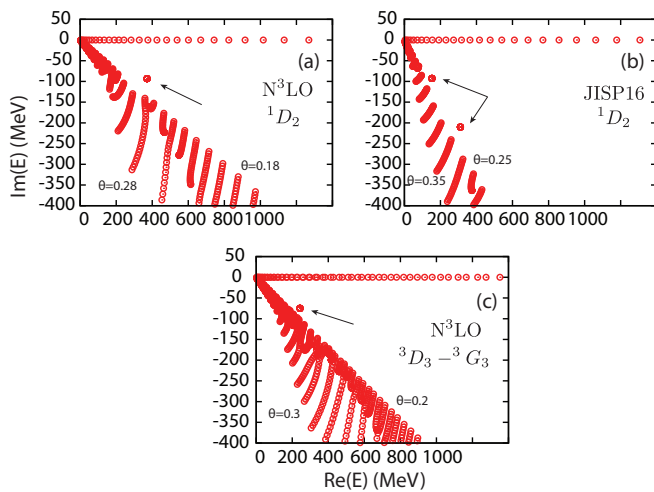} 
  \caption[T]{\label{Fig3}
  (Color online) Spectrum of the $^1$D$_2$ channel with the N$^3$LO (a) and JISP16 (b) interactions. Notice the two poles in the case of JISP16.
  In panel (c) it is the coupled $^3$D$_3$-$^3$F$_3$ spectrum for the N$^3$LO interaction. }
\end{figure}
are both fitted to higher energies than N$^2$LO$_{opt}$. 
The $^1$D$_2$ channel shown in Fig. \ref{Fig3} presents an interesting case since both interactions
support resonant states and, in the case of JISP16, we find two resonances. 
According to our calculations the N$^3$LO resonant position is
at (371.111 -i93.169) MeV. We notice that at an energy of about 400 MeV for the real part and about -150 MeV for the imaginary part,
there is also a state that shows a stabilization pattern with respect to the rotation parameter. This state, however, is not as stable as the other
cases we examine in this work, so we do not investigate it further.
For the JISP16 interaction we observe  clearly two resonant positions in the $^1$D$_2$ channel :
(153.155 -i92.725) MeV and a  broader structure at (311.008 -i210.065) MeV (see panels (a) and (b) of Fig. \ref{Fig3}).

In panel (c) of Fig.\ref{Fig3} we show the spectrum for the coupled $^3$D$_3$-$^3$G$_3$ channel for the N$^3$LO potential which supports a
resonant state at a position: (219.218 -i75.162) MeV. For the same channel, not shown here, the JISP16 interaction supports a broader 
structure at (200.084  -i115.138) MeV. In this JISP16 case the rotation angle was varied from $\theta$ = 0.32 to 0.42 rad in order to reveal the broader resonant position.

\begin{figure}[h!] 
  \includegraphics[width=\columnwidth]{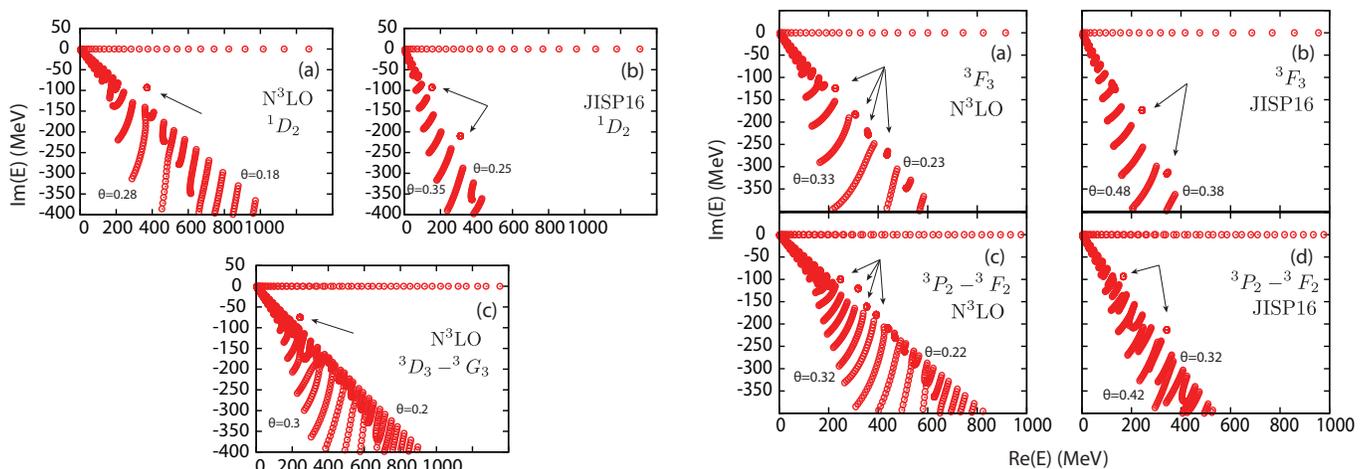} 
  \caption[T]{\label{Fig4}
  (Color online) Same as in Fig. \ref{Fig3} but for the $^3$F$_3$ channel (panels (a),(b)) and $^3$P$_2$-$^3$F$_2$ (panels (c),(d)) for the 
  N$^3$LO and JISP16 interactions respectively. See also Table \ref{Tab:2}.}
\end{figure}

In Fig. \ref{Fig4} we show our results for the uncoupled $^3$F$_3$ and the coupled $^3$P$_2$-$^3$F$_2$ channels using the N$^3$LO and JISP16
forces. For these partial waves we find multiple resonances supported by the interactions. The chiral potential supports at least four
resonant-like structures for the $^3$F$_3$ channel and for the coupled  $^3$P$_2$-$^3$F$_2$ channel, 
where in the latter case we also see some indications for 
resonant structures at energies larger than 400 MeV CoM. According to our calculations, for the JISP16 interaction we find two resonant-like structures for each of the
$^3$F$_3$ and $^3$P$_2$-$^3$F$_2$ channels. 
\begin{table}[ht]
\caption{Resonant parameters that correspond to the arrows of  Fig. \ref{Fig4}, computed with the $\theta$-trajectory 
method for the uncoupled  $^3$F$_3$ and  coupled  $^3$P$_2$-$^3$F$_2$ channels, using the N$^3$LO and JISP16 interactions.  }
\begin{tabular}{cccc}
\hline
 \multicolumn{2}{c}{N$^3$LO}&\multicolumn{2}{c}{JISP16}\\
 $^3$F$_3$ &  $^3$P$_2$-$^3$F$_2$  &$^3$F$_3$&  $^3$P$_2$-$^3$F$_2$  \\
\hline\hline
(227.0  -i124.3)&  (247.8  -i99.8)   & (242.3 -i173.4)  &  (161.2 -i95.2)\\
(307.2  -i182.5)&  (318.9 -i120.5)  & (342.6  -i315.6) &  (342.4 -i213.2)\\
(360.5   -i227.9)& (353.5  -i160.6) &  &  \\
(435.3 -i273.2)&  (392.0  -i179.1)  & & \\
\hline
\end{tabular}
\label{Tab:2}
\end{table}
After applying the $\theta$-trajectory stabilization method we gather the results in Table \ref{Tab:2}.
We notice that  the states are broader than the structures we find for the $^3$P$_{0}$ and $^3$P$_1$ channels.
The positions of the resonant states in each channel depend on the interaction we use. In this sense, the situation is distinctly different from the deuteron ground state that all the realistic NN
interactions are constrained to describe with high accuracy.
Of course, this dependence of the scattering resonances on the NN interaction may be expected since we are investigating a kinematic region which is sensitive to short range nuclear physics, a sector that each NN interaction models differently. 

To show that the resonance locations are fixed by NN model assumptions and not by the couplings of low and high momenta alone, we made a test which demonstrates that the $^3$P$_1$ resonant state for the N$^3$LO potential is invariant under 
similarity renormalization group (SRG) transformations \cite{bogner}, down to a
scale of $\lambda$ = 1.5fm$^{-1}$.  A scale of $\lambda$ = 1.5fm$^{-1}$, taken as a maximum relative momentum,
corresponds roughly to an energy of 187 MeV in the laboratory frame.
The real part of energy of the $^3$P$_1$ resonant-like state (see Table \ref{Tab:1}) is above the energy implied by this SRG scale (i.e. E$_{res}^{lab}$ $\sim$ 2$\cdot$214 = 428 MeV). Intuitively, one would think that the NN physics up to this SRG scale will remain invariant but above this scale one might expect some changes so we are motivated to test for SRG scale invariance using this resonance as the test case.

\begin{figure}[h!] 
  \includegraphics[width=\columnwidth]{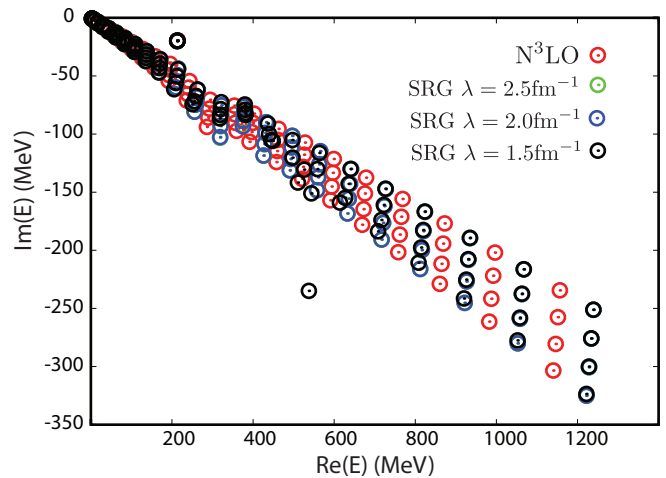} 
  \caption[T]{\label{Fig5}
  (Color online) $^3$P$_1$ channel CS spectrum for the bare N$^3$LO force and for three different SRG evolution scales. 
  The resonant-like structure remains
  invariant. The green circle that corresponds to SRG $\lambda$ = 2.5 fm$^{-1}$ practically overlaps with the results of the other SRG evolution scales and it is not visible. A solution is observed that corresponds to a $\lambda$ = 1.5 fm$^{-1}$ with an imaginary part of about -250 MeV. This is a non-resonant continuum state which happened to depart from the approximate 2$\theta$ line.}
\end{figure}

Using the N$^3$LO interaction in the $^3$P$_1$ channel we performed three SRG transformations,
that reduce the couplings of low momentum to high momentum states and change the off-shell components of the potential, leaving the on-shell properties invariant.
In Fig.\ref{Fig5} we observe that the position of the $^3$P$_1$ resonant-like state is invariant under these SRG evolutions, for the SRG transformation scale as low as $\lambda$ = 1.5fm$^{-1}$ starting from a bare N$^3$LO potential. 
For these calculations the rotation angle was rotated from $\theta$ = 0.1 rad to 0.2 rad with a step of 0.05 rad. 

\textit {Conclusions.} In this work we applied the CS method to study resonant features of the NN interaction in energy regions above the usual 300 to 350 MeV laboratory energy.  While this region exceeds the energies for which these non-relativistic
interactions are developed, we obtain resonances motivating further exploration and we
demonstrate the robust characteristics of our techniques.

We analyzed the resonant features that we found in several channels and we studied the stability of the results using 
the $\theta$-trajectory method whose validity has been 
demonstrated for local and schematic potentials. In our work we find similar convergence patterns for resonant states to what other authors have found. 
Among the numerous NN resonances that we found, the one in the $^3$P$_1$ channel appeared  close to the real energy axis. 

In our formalism, the widths of the states should be viewed as total decay widths, and since we did not consider special decay channels in our analysis, we cannot say if the $np$ system will decay by either emitting mesons or baryon resonances. 
Our numerical results show that the positions of these states in the complex
plane depend strongly on the form of the underlying interaction.  
Hence these NN-dependent S-matrix poles are distinguished from the deuteron pole 
whose properties are shared by all realistic NN interactions to high accuracy.
 
We focused on a specific force in one channel, in particular the chiral N$^3$LO in the $^3$P$_1$ channel,
to demonstrate that the position of the resonant-like state is invariant under the action of SRG transformations.

Finally, it worth mentioning that analysis of NN scattering data has not, to our knowledge, reported resonant poles for the $^3$P$_0$ and $^3$P$_1$  partial waves, but rather for the
$^1$D$_2$,  $^3$F$_3$, $^3$P$_2$-$^3$F$_2$ and recently for the $^3$D$_3$-$^3$G$_3$. 

Among our achievements, we have demonstrated the stability of the CS method for identifying resonant
states. This demonstration helps support the adoption of these techniques for investigating resonances in finite nuclei
where we anticipate the interest will be in low-energy applications. 

\textit{Acknowledgements.} We gratefully acknowledge valuable discussions with I. I. Strakovsky.
This work was supported by the US DOE under grants No. DESC0008485 (SciDAC/NUCLEI) and DE-FG02-87ER40371.

\bibliography{CS_real2}    

\end{document}